\documentclass[a4paper,11pt]{article}
\usepackage{pos}
\usepackage{amsmath}

\title{Ending inflation with a bang: Higgs vacuum decay in $R^2$ gravity}

\author*[a, b]{Andreas Mantziris}

\affiliation[a]{Department of Physics, Imperial College London, London, SW7 2AZ, United Kingdom}

\affiliation[b]{Faculty of Physics, University of Warsaw, ul. Pasteura 5, 02-093 Warsaw, Poland}

\emailAdd{andreas.mantziris@fuw.edu.pl}

\abstract{According to the current experimental data, the Higgs vacuum appears to be metastable due to the development of a second lower ground state in its potential. Consequently, vacuum decay would induce the nucleation of true vacuum bubbles with catastrophic consequences for our Universe and therefore we are motivated to study possible stabilising mechanisms in the early universe. In our latest investigation (2207.00696), we studied the electroweak metastability in the context of the observationally favoured model of Starobinsky inflation. Following the motivation and techniques from our first study (2011.037633), we obtained constraints on the Higgs curvature coupling $\xi$, while embedding the SM on the modified gravity scenario $R+R^2$, which introduces Starobinsky inflation naturally. This had significant repercussions for the effective Higgs potential in the form of additional negative terms that destabilize the false vacuum. Another important aspect lay in the definition for the end of inflation, as bubble nucleation is most prominent during its very last moments. Our results dictated that these stronger lower $\xi$-bounds are very sensitive to the final moments of inflation, where spacetime deviates increasingly from de Sitter.}

\FullConference{%
  41st International Conference on High Energy physics - ICHEP2022\\
  6-13 July, 2022\\
  Bologna, Italy
}

\newcommand{\Lagr}{\mathcal{L}}
\def\Nbub{\langle {\cal N}\rangle}

\renewcommand{\logo}{\relax}

\begin{document}
\maketitle

\section{Introduction}

\subsection{The electroweak vacuum metastability}

The Standard Model (SM) of particle physics has famously been one of our most successful theories, with remarkable accuracy in its predictions about the behaviour and interactions of the subatomic particles of our Universe. Even though there are various shortcomings of the theory that would require physics beyond the SM \cite{Berti:2022wzk}, in principle from the SM's internal point of view, it remains valid up to very high energy scales, beyond the TeV's we probe with accelerators. Therefore, in this work we adopt a conservative approach that utilises the SM with the addition of cosmological inflation as a minimal model to describe the early Universe. 

\begin{figure}[h]
\centering
\includegraphics[width=0.92\linewidth]{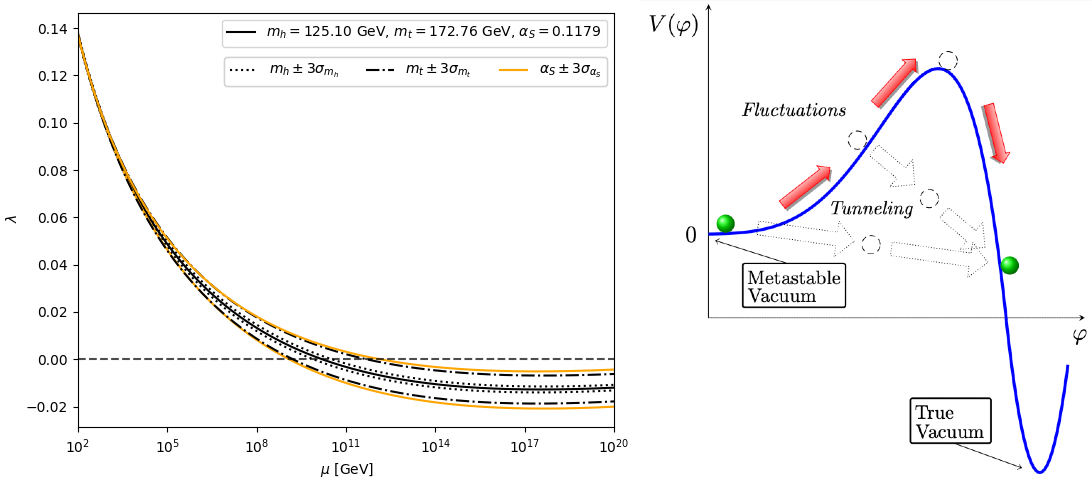}
\caption{Left: The evolution of the Higgs self interaction $\lambda$ with the renormalisation scale $\mu$ for the current experimental measurements of the Higgs $m_h$ and top quark $m_t$ masses, and the strong coupling $a_S$ to $3 \sigma$ variance \cite{Mantziris:2022fuu}. Right: A metastable double-well potential for a scalar field $\varphi$ that can decay to its true vacuum via fluctuations, quantum tunnelling, or a combination of both \cite{Markkanen:2018pdo}.}
\label{fig:lambda-double_well}
\end{figure}

According to the experimental measurements of SM parameters \cite{ParticleDataGroup:2020ssz}, the electroweak (EW) vacuum that the Higgs field currently resides in, is prone to decay to a lower ground state. This additional vacuum state in the Higgs potential, called the \textit{true} vacuum, and the potential barrier that separates it from the \textit{false} EW vacuum come from the running of the Higgs quartic interaction $\lambda$ that switches sign at $ \mu \sim 10^{10}$ GeV, as shown in  Fig. \ref{fig:lambda-double_well}. The scale dependence of the Higgs self coupling is calculated according to its $\beta$-function, where the competing bosonic and fermionic contributions are dominated by their heaviest particles, the Higgs and the top quark, weighted by their corresponding masses $m_h, \, m_t$.

Vacuum decay can proceed either via thermal fluctuations over the barrier, tunnelling through it, or a mixture of the two. Since the EW vacuum has survived throughout our cosmological history, we can constrain fundamental physics so that they allow for this long-lasting metastability. The potential of the Higgs field $h$ at tree-level on a curved background is given by
\begin{align}
  V_{\rm H}(h, \mu, R) = \frac{\xi(\mu)}{2}R h ^2 + \frac{\lambda(\mu)}{4} h^4 \, ,
  \label{eq:Vtree}
\end{align}
where $\xi$ is the non-minimal coupling between the field and spacetime curvature $R$. This Higgs curvature coupling is the last unknown renormalisable SM parameter, because it cannot be probed experimentally in the flat spacetime of the late Universe. Thus, demanding the stability of the EW vacuum in the cosmological context constrains $\xi$ accordingly, so that vacuum decay is suppressed.

\subsection{Bubble nucleation from vacuum decay}

Vacuum decay takes place at a point in spacetime and subsequently excites its surroundings in a chain reaction, that results in a bubble of true vacuum growing with relativistic velocity. Since our Universe is still in the metastable vacuum, no bubbles were nucleated during our cosmological evolution. Therefore, the expectation value for the number of true vacuum bubbles $\mathcal{N}$ has to obey 
 \begin{align}
        \Nbub \lesssim 1 \, ,
        \label{eq:constraint}
    \end{align}
in order to be compatible with observations \cite{Markkanen:2018pdo}, and it is given by the product of the decay rate $\Gamma$ per spacetime volume $\mathcal{V}$ integrated over the past lightcone, where $g$ is the determinant of the metric,
 \begin{align}
    d\Nbub = \Gamma d{\cal V} \, \Rightarrow \, \, \Nbub = \int_{\substack{ \rm past }} d^4x \sqrt{-g} \Gamma(x) \, .
    \label{eq:Nbubs}
    \end{align}

In the context of the SM, the bubble interior is a singularity, but its exact description depends on the shape of the potential, whether the true vacuum is bounded from below or not, and the UV completion of the theory, where quantum gravity effects become important. However, these considerations are beyond our scope, because we are interested in the number of bubbles in our false vacuum Universe and not in the exotic physics within them, since even one nucleation event is ruled out, as the bubble would expand and consume our Universe in a rapid and violent manner.

Despite the decay rate being slow in the late Universe, it could have been enhanced significantly at early times, and thus we are motivated to study vacuum decay during the period of cosmological inflation. This corresponds to an epoch of exponential expansion of the Universe, whose duration we quantify with the number of $e$-foldings of inflation $N= \mathrm{ln} \left( \frac{a_{\rm inf}}{a}\right)$, where $a$ is the scale factor and the index ``inf'' denotes the end of inflation. We integrate Eq. (\ref{eq:Nbubs}) backwards in time from $N_{\rm inf} =0$ until the total duration $N_{\rm start}$, which is bounded from below according to observations at $N_{\rm start} \gtrsim 60$, 
\begin{align}
    \Nbub =  \frac{4\pi}{3}\int_0^{N_{\mathrm{start}}} dN \left( \frac{a_{\mathrm{inf}} \left(\eta_0-\eta\left(N\right)\right)}{e^{N}} \right)^3 \frac{\Gamma(N)}{H(N) } \, ,
    \label{eq:Nbub}
\end{align}
where $H=\frac{\dot{a}}{a}$ is the Hubble rate, and $\eta_0 - \eta(N)$ the comoving radius of the spacetime volume from the present day to $N$ $e$-folds before the inflationary finale.

The classical solutions for the transition from false to true vacuum are called instantons. For the high Hubble scales during inflation, the Hawking-Moss (HM) process dominates the decay rate
\begin{align}
	\Gamma \approx \left(\frac{R}{12}\right)^2 e^{-\frac{384 \pi^2  \Delta V_{\rm H} }{R^2}} \, ,
\end{align} 
where $ \Delta V_{\rm H}$ corresponds to the barrier height, since in the HM regime the field ``goes'' over the barrier \cite{Hawking:1981fz}. Therefore, there are two parallel streams of computations that start independently, before coalescing into a complete calculation that constrains the non-minimal coupling. On the one hand, the particle physics aspect involves $\Gamma$ via the estimation of the effective Higgs potential in curved spacetime, while on the other hand, there are cosmological quantities associated with the lightcone volume, that are subject to the choice of the inflationary potential $ V_{\rm I}(\phi)$, the total duration and the endpoint of inflation. After inserting them in Eq. (\ref{eq:Nbubs}) and imposing the condition (\ref{eq:constraint}), while assuming that inflation lasts for 60 $e$-foldings, we finally obtain the bounds $\xi \geq \xi_{\Nbub=1}$ in addition to cosmological implications from the time of predominant bubble nucleation.

\section{The effective Higgs potential in Starobinsky inflation}

Beyond the tree-level potential (\ref{eq:Vtree}), we include Minkowski terms to 3-loops and curvature corrections in de Sitter (dS) at 1-loop, and a small gravitational correction is radiatively generated,
\begin{align}
       V_{\rm H}(h, \mu, R) = \frac{\xi}{2} R h^2 + \frac{\lambda}{4} h^4 + \frac{\alpha}{144} R^2 + \Delta V_{\rm loops} \, ,
 \end{align}
where the loop contribution from all the SM degrees of freedom in dS reads
\begin{align}
     \Delta V_{\rm loops} = \frac{1}{64\pi^2} \sum\limits_{i=1}^{31}\bigg\{ n_i\mathcal{M}_i^4 \bigg[\log\left(\frac{|\mathcal{M}_i^2 |}{\mu^2}\right) - d_i \bigg] +\frac{n'_i R^2}{144}\log\left(\frac{|\mathcal{M}_i^2 |}{\mu^2}\right)\bigg\}  \, ,
\end{align}
where $\mathcal{M}_i$ is each particle's effective mass, and $n_i, \, d_i, \, n_i'$ are constant numbers \cite{Markkanen:2018bfx}. The ``unphysical'' $\mu$-dependence of the potential can be eliminated through Renormalisation Group Improvement (RGI), where the scale is chosen as $\mu=\mu_*(h,R)$ so that $\Delta V_{\rm loops} (h, \mu_*, R) = 0 \,  $, which implies a valid perturbative expansion. Thus, the RGI effective Higgs potential is written as
\begin{align}
    V_{\rm H}^{\rm RGI}(h, R) = \frac{\xi(\mu_*(h, R))}{2}  R h ^2 +\frac{\lambda(\mu_*(h, R))}{4}  h^4 + \frac{\alpha(\mu_*(h,R))}{144} R^2 \, .
    \label{eq:Veff}
\end{align}

\begin{figure}[h]
    \centering
\includegraphics[width=0.79\linewidth]{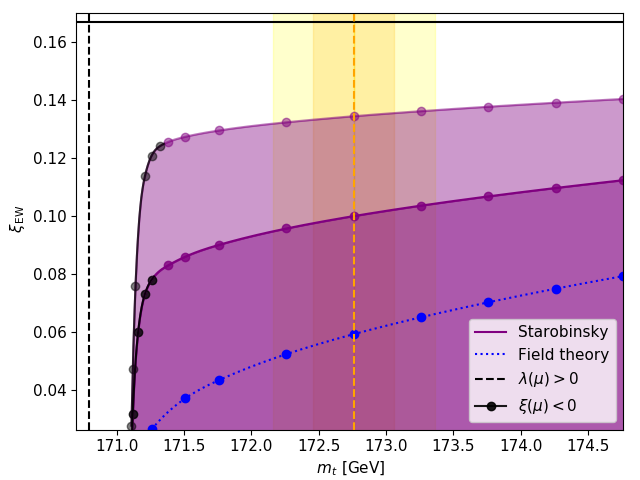}
\caption{Lower bounds on the non-minimal coupling's electroweak value $\xi_{\rm EW}$ with respect to the top quark mass $m_t$ around its central value (vertical dashed orange line) with $\pm \sigma_{m_t}$ and $\pm 2\sigma_{m_t}$ (orange shades). The purple areas denote the excluded parameter space when the end of inflation is set at $\dot{H}/H^2 = -1/4$ (darker) and $\dot{H}/H^2 = -1$ (lighter). The blackened curves show the threshold below which the false vacuum is pushed to higher field values due to $\xi$ turning negative as it runs. The results from the field theory case of \cite{Mantziris:2020rzh} are shown by the dotted blue curve for comparison. The  horizontal black line lies at the conformal point $\xi = 1/6$ and the vertical dashed black line depicts the minimum $m_t$, below which the EW vacuum is stable. \cite{Mantziris:2022fuu} }
\label{fig:bounds}
\end{figure}

However, when embedding the SM in $R+R^2$ gravity, which gives rise to Starobinsky inflation \cite{Starobinsky:1979ty}, the calculation of the effective potential is more complicated. Starting from the action of the Higgs field in the Jordan frame, where the geometric $R^2$-term is added to the Einstein-Hilbert $R$-term of General Relativity (GR), 
\begin{align}
    S = \int d^4x \sqrt{-g_J} \left[ \frac{M_P^2}{2} \left( 1 - \frac{\xi h^2}{M_P^2} \right) R_J + \frac{1}{12 M^2}R_J^2 +  \frac{1}{2} g_J^{\mu \nu} \partial_{\mu} h\partial_{\nu} h  - \frac{\lambda}{4}h^4 \right] \, ,
\end{align}
where $M_P = 2.435 \times 10^{18}$ GeV is the reduced Planck mass and $M$ is a a small dimensionless parameter. We perform the necessary conformal transformation to the Einstein frame $g_{J \, \mu \nu} \rightarrow g_{\mu \nu}$, where we recover GR with the addition of a scalar field, the inflaton $\phi$, in the matter sector. After a field redefinition $h \rightarrow \Tilde{h}$ that allows us to RG improve the potential in the dS limit, where $\phi$ is approximately constant, we have to perform two more field redefinitions $\phi \rightarrow \Tilde{\phi}$ and $ \Tilde{h} \rightarrow \rho$, in order to express the Lagrangian in a canonical form
\begin{align}
	\Lagr \approx  \frac{M_P^2}{2}R + \frac{1}{2}\partial_{\mu} \Tilde{\phi} \partial^{\mu} \Tilde{\phi} + \frac{1}{2} \partial_{\mu} \rho \partial^{\mu}  \rho   - \Tilde{U}(\Tilde{\phi}, \rho) \, .
\end{align}
The characteristic Starobinsky potential $V_{\rm I}(\Tilde{\phi}) = \frac{3 M^2 M_P^4}{4} \left(1-e^{-\sqrt{\frac{2}{3}}\frac{\Tilde{\phi}}{M_P}} \right)^2$ has been generated ``naturally'' in the full potential of the theory,
\begin{align}
\Tilde{U}(\Tilde{\phi}, \rho) = V_{\rm I}(\Tilde{\phi}) +  m_{\rm eff}^2 (\Tilde{\phi}, \mu_*) \frac{\rho^2}{2} + \lambda_{\rm eff} (\Tilde{\phi}, \mu_*) \frac{\rho^4}{4} + \frac{\alpha (\mu_*)}{144} R^2 (\Tilde{\phi}) + \mathcal{O}(\rho^6/M_P^2) \, ,
\end{align}
where the effective Higgs potential resembles Eq.(\ref{eq:Veff}) with the addition of a Planck suppressed $\rho^6$-term and the $\Tilde{\phi}$-dependent  contributions to the effective mass and self-coupling 
\begin{align}
    m_{\rm eff}^2 &= \xi R + 3 M^2 M_P^2 \Xi \left(1-e^{-\sqrt{\frac{2}{3}}\frac{\Tilde{\phi}}{M_P}}\right) e^{-\sqrt{\frac{2}{3}}\frac{\Tilde{\phi}}{M_P}} + \frac{\Xi}{M_P^2} \partial_{\mu} \Tilde{\phi} \partial^{\mu} \Tilde{\phi} \, , \\
    \lambda_{\rm eff} &= \lambda + 3 M^2\Xi^2 e^{-2\sqrt{\frac{2}{3}}\frac{\Tilde{\phi}}{M_P}} + \frac{4 \left[ \xi R + \Delta m^2_1 \right] \Xi^2}{M_P^2} +  \frac{4 \Xi^3}{M_P^4} \partial_{\mu} \Tilde{\phi} \partial^{\mu} \Tilde{\phi} \,,
\end{align}
respectively, where $\Xi (\mu_*) = \xi (\mu_*)  -\frac{1}{6}$. Since $0<\xi_{\rm EW}<1/6$ in the HM regime, these additional negative terms, in particular the quadratic ones, destabilise the vacuum. This means that stronger $\xi$-bounds are needed, compared to the field theory case of \cite{Mantziris:2020rzh}, to counter them and allow for the survival of the EW vacuum, as shown in Fig. \ref{fig:bounds}.

Therefore, in Starobinsky inflation $\xi_{\rm EW} \gtrsim 0.13$, as opposed to the field theory result $\xi_{\rm EW} \gtrsim 0.06$, for $m_t = 172.76$ GeV and where the end of inflation is set at $\ddot{a}=0$. However, because bubbles are predominantly produced very close to the inflationary finale, due to the negative terms in $m^2_{\rm eff}$, we have to be more conservative with our definition of inflation's endpoint, for the $\xi$-bounds to be valid. This is due to the dS approximations used when calculating the effective potential and $\Gamma$, but spacetime deviates increasingly from dS towards the end of inflation. Thus, if inflation lasts until $\frac{\ddot{a}}{a} = \frac{H^2}{2}$, we obtain the more trustworthy constraint $\xi_{\rm EW} \gtrsim 0.1$, shown in dark purple in Fig. \ref{fig:bounds}.

\section{Conclusions}
In this work, we presented an overview of our latest study \cite{Mantziris:2022fuu}, where we have assumed a minimal model to describe the early Universe, consisting of the SM and Starobinsky inflation. The latter is supported by observations and arises from the simple modification of gravity $R+R^2$, where a higher-order geometric term is included in the action. Utilising the techniques and software from \cite{Mantziris:2020rzh}, we obtained stronger vacuum decay constraints on the Higgs-curvature coupling 
        \begin{align}
            \xi_{\rm EW} \gtrsim 0.1 > 0.06  \, ,
        \end{align}
with a state-of-the-art RG improved effective Higgs potential, where the negative terms that have arisen are in competition with the $\xi R$ factor. Since bubble nucleation takes place in the last moments of inflation, the dS approximations for $\Gamma_{\rm HM}$ start to break down, and therefore it is necessary to consider the dynamics of reheating, in order to acquire more precise and definite constraints on $\xi$. However, the early time behaviour of the potential is identical to the field theory case, and thus, we obtain the same hints against eternal inflation as in \cite{Mantziris:2020rzh}.

\acknowledgments
The author is grateful for the STFC PhD studentship and the supervision and collaboration of A. Rajantie and T. Markkanen on \cite{Mantziris:2020rzh}, which was the foundation for the parallel talk at ICHEP2022.

\bibliographystyle{JHEP}
\bibliography{references.bib}

\end{document}